\documentclass[12pt]{article}
\usepackage{epsfig,axodraw}

\begin{document}

\title{Code for prompt numerical
computation of the leading order GPD evolution}
\author{A.V.~Vinnikov\footnote{E-mail: vinnikov@theor.jinr.ru}\\
{\it DESY, 15738 Zeuthen, Germany and JINR, 141980 Dubna, Russia} }
\maketitle
\begin{abstract}
This paper describes the design and work of a set of computer routines
capable for numerical computation of generalized parton distributions
(GPDs) evolution at the leading order. The main intention of this work
is to present a fast-working computer code making possible
fitting of GPDs parameters to the data on hard electron-nucleon
scattering.
\end{abstract}

\section{Introduction}
\begin{figure}[here]
\centering
\begin{picture}(150,110)(0,-5)
\ArrowLine(40,25)(50,85)
\ArrowLine(100,85)(110,25)
\SetWidth{2.0}
\ArrowLine(5,15)(40,25)
\ArrowLine(110,25)(145,15)
\Text(5,70)[]{{\large $(x+\xi)\frac{p_1+p_2}{2}$}}
\Text(145,70)[]{{\large $(x-\xi)\frac{p_1+p_2}{2}$}}
\Text(5,1)[]{{\large $p_1$}}
\Text(145,1)[]{{\large $p_2$}}
\COval(75,25)(10,35)(0){Gray}{Gray}
\end{picture}
\caption{Partonic picture for GPDs.
The value of $\xi$ is given by $\xi=x_B/(2-x_B)$, where $x_B$ is Bjorken
variable.}
\label{kinemat1}
\end{figure}
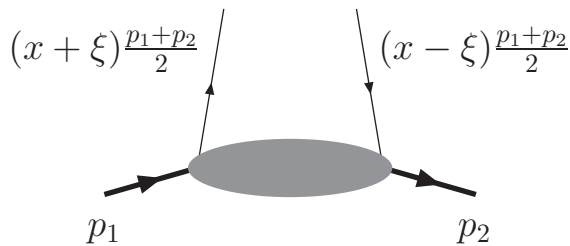

Generalized parton distributions \cite{gpds1,gpds2,gpds3,gpds4,gpds5}
parameterize matrix elements of quark and gluon twist-2 operators
between nucleon states with non-equal momenta.
They can be represented as functions of fractions $x$ and $\xi$
of longitudinal momentum
of the proton carried by the parton and of $t=(p_1 - p_2)^2$,
the square of the 4-momentum transfer between initial and final
protons\footnote{The variable $t$ does not enter
the evolution equations and is therefore omitted in this paper.}
(see. Fig.~\ref{kinemat1}).

GPDs are the basis for description of hard elastic lepton-nucleon
scattering in terms of QCD, just like the ordinary PDFs are the
basis of QCD description of the DIS. Besides the description
of hard elastic reactions, GPDs give
access to the orbital momentum of partons in the nucleon \cite{jirule}
and 3-dimensional nucleon structure \cite{impact1,impact3,impact4,impact2}.
The importance of GPDs for understanding of the nucleon structure
raised a problem of their extraction from hard elastic scattering data,
first of all from deeply virtual Compton scattering (DVCS).
The activity of measuring DVCS observables at lepton-nucleon facilities
is quite intense
\cite{h1dvcs,zeusdvcs,hermesdvcs,clasdvcs1,clasdvcs2}.

Similar to PDFs, GPDs are subject of perturbative QCD evolution
which should be taken care of in extraction of GPDs from the data
on lepton-nucleon scattering. The LO kernels for GPDs evolution are
well-known \cite{gpds1,gpds2,gpds3,gpds4,gpds5,rad_pol,bluevol}.
The NLO results have been published \cite{belevol}.
At present, there exists a code \cite{mk_freund1,mk_freund}
which can be used for leading and next-to-leading order evolution of GPDs.
However, on a 1.7 GHz Pentium-4 machine this code takes about
15 seconds for LO evolution and about 40 seconds for the NLO which
is not fast enough for performing fitting procedure
requiring very large number of evolution routine calls.
Evolution equations for GPDs at the LO can be also solved by a decomposition
of GPDs in series of special functions \cite{manashov1,manashov2}.
The advantage of this method is that it does not require introduction
of a difference scheme for $Q^2$. However, the decomposition is not
trivial for numerical implementation and the number of terms in the series
required to achieve a reliable precision can be large (depending on the input GPDs).

In the present paper another code is introduced\footnote{The code
is available at http://www.ifh.de/\~{}vinnikov/gpd\_evol.tar.gz},
which does the LO GPD evolution with a reasonable precision
for $\sim 10^{-3}$ seconds, making possible its usage for fitting.
The fast work was achieved by optimization of the computational
procedure. It uses 4-th order Runge-Kutta method for solving the
evolution equations. Since in the broad DGLAP region ($|x|>\xi$)
the variation of GPDs is not sizable, and in the narrow
ERBL region ($|x|<\xi $) is quite significant, the convolution integrals
are performed on a logarithmic grid so that the number
of points in ERBL and DGLAP regions is the same.
The code itself is also optimized. For example,
the operation of division is avoided as much as possible. This was done
since division on the machines commonly used (such as Pentium-IV)
is performed 6 times
slower than multiplication. So, if an expression to be computed
reads $a/b/c$, its optimized version is $a/(b*c)$.
Since the NLO expressions are considerably larger then the LO ones,
their implementation is not done at the moment.

\section{Evolution kernels}
The LO kernels for GPDs evolution are well-known. In the code
those published in \cite{gpds5,rad_pol,bluevol}
are used\footnote{Literally, the expressions from Ref.~\cite{bluevol}
were taken. Due to the symmetry relations of singlet GPDs, these kernels are
equivalent to those published in Ref.~\cite{gpds5}, (see Eq.~5.21
of Ref.~\cite{gpds5}).}.
The numerical computation of the evolution requires
that singularities in the kernel to be cancelled analytically.
The kernels with regularized singularities in the
form they take in the code are given in the Appendix.

The kernels assume the definition of the GPDs given in \cite{markus_rep}.
In the forward limit they obey
\begin{eqnarray}
H_{NS}(x,\xi=0,t=0) = u(x) - d(x) \, , \nonumber \\
H_{Q}(x,\xi=0,t=0) = u(x)+ d(x) + s(x)  \, , \nonumber \\
H_G(x,\xi=0,t=0) = xg(x) \, , \\
\widetilde H_{Q}(x,\xi=0,t=0) = \Delta u(x)
+ \Delta d(x) +\Delta s (x)  \, , \nonumber \\
\widetilde H_G(x,\xi=0,t=0) = x\Delta g(x) \, , \nonumber
\end{eqnarray}
with $q(x) = q_V(x) + \bar q(x)$, $\Delta q(x) = \Delta q_V(x) +
\Delta \bar q(x)$, 
$\bar q (-x) = -\bar q(x)$, $g(-x) = -g(x)$,
$\Delta \bar q (-x) = \Delta \bar q(x)$, $\Delta g(-x) = \Delta g(x)$.
Note that quark singlet and non-singlet densities are not divided
by $n_f$. This is important in the singlet sector, where non-diagonal
parts of the kernel depend on the normalization of the quark singlet
distribution. For non-singlet channel, isovector combination was chosen
to represent the work of the code. It is clear that any
other non-singlet combination of GPDs can be taken as well.

In the singlet sector, the symmetry properties $H_Q(-x,\xi,t)=
-H_Q(x,\xi,t)$, $H_G(-x,\xi,t)= H_G(-x,\xi,t)$,
$\widetilde H_Q(-x,\xi,t)= \widetilde H_Q(x,\xi,t)$ and
$\widetilde H_G(-x,\xi,t)=-\widetilde H_G(x,\xi,t)$ are used.
This allowed to reduce
the computational time by a factor of 4. Thus, the singlet
kernels are given only for $x \ge 0$.

For all parts of the kernel, convolution at $x=\pm\xi$ was done separately.
The kernels at $x=\pm \xi$ were taken as $x\to \xi$ limits of the corresponding
DGLAP or (equivalently for $x\to \pm \xi$) ERBL expressions.

The running strong coupling constant is taken as a solution of the NLO
RG equation
\begin{equation}
\frac {d\, \alpha_s (\mu)}{d\, \ln(\mu^2)} = 
-b_0 \alpha_s^2(\mu) - b_1\alpha_s^3(\mu) -b_2\alpha_s^4(\mu)
\label{run_coup}
\end{equation}
where the coefficients $b_0$ and $b_1$ are given by
\begin{eqnarray}
b_0 &=& \frac{1}{12\pi}\left ( 33 - 2 N_f \right ) , \nonumber \\
b_1 &=& \frac{1}{24\pi^2}
\left ( 153 - 19 N_f \right )\, , \\
b_2 &=& \frac{1}{3456\pi^3}
\left ( 77139 - 15099 N_f + 325 N_f^2\right )\, . \nonumber
\end{eqnarray}
Here $N_F$ is the number of active quark flavors.
The initial condition for Eq. \ref{run_coup} is taken as the value
of $\alpha_s(M_Z)$. Then the equation is solved back to smaller values
of $\mu$. Down to $\mu=m(b) = 4.3$ GeV the number of active flavors is taken
to be equal 5. Between $\mu = m(c) =1.3$ GeV and $\mu = m(b)$,
$N_f=4$ is used. Below $\mu=m(c)$ $N_f$ is 3. Note that the number of
active flavors $N_f$ entering the $\alpha_s(\mu)$ RG equation
is different from the number of flavors in the proton $n_f$ which enters 
into the quark-gluon evolution kernel. In the code, $n_f$ is always 
equal to 3.

\section{Computation of the convolution integrals}
The integrals are computed on a logarithmic grid.
The required parameterization for non-singlet case reads:
\begin{eqnarray}
x_i = -\delta e^{-(i-2n)} + \delta \quad (-1\le x\le 0, \, 0\le i \le 2n)\, ,
\\
x_i = \delta e^{(i-2n)} - \delta \quad (0< x\le 1, \, 2n< i \le 4n)\, ,
\label{lngrid}
\end{eqnarray}
where $\gamma$ and $\delta$ are given by
\begin{equation}
\gamma=\frac{1}{n}\ln(\frac{1}{2} + \frac{1}{2\xi}\sqrt{9\xi^2-12\xi+4})\, ,
\quad \delta = \frac{\xi^2}{1-2\xi} \, .
\label{lngridpar}
\end{equation}
This grid has the following useful properties:
\begin{equation}
x_0 = -1,\, x_{n} = -\xi,\, x_{2n} = 0, x_{3n} = \xi, x_{4n} = 1 \, 
\end{equation}
i.e. it provides the same level of detalization for ERBL and DGLAP regions.

For the singlet case the grid size is reduced so that to cover only
$0\le x\le 1$ interval:
\begin{equation}
x_i = \delta e^{\gamma i} - \delta \quad ( 0\le i \le 2n)\, ,
\end{equation}
so that
\begin{equation}
x_0 = 0,\, x_{n} = \xi,\, x_{2n} = 1\, .
\end{equation}
The parameters $\gamma$ and $\delta$ are the same as in the non-singlet case
(Eq.~\ref{lngridpar}).

After introduction of the logarithmic grid
the convolution integrals can be computed using the standard
equidistant Simpson's formula taking into account the Jacobian
for the grid:
\begin{equation}
\int\limits_{-1}^1 f(x) dx = \int\limits_{0}^{2n}\gamma(\delta-x_i)f(x_i) di
+\int\limits_{2n}^{4n}\gamma(\delta+x_i)f(x_i) di\, .
\end{equation}
The $\frac{0}{0}$ uncertainty in the integrals can be resolved with the
L'Hopital's rule. Since for the numerical integration the Simpson's
$\left ( {\cal O}(\Delta x^4) \right ) $ method is used, the numerical
computation of the derivative requires ${\cal O}(\Delta x^3) $ precision.
Depending on the position of the singularity (availability of other
grid points on the left and on the right), one of the following
formulas is be used:
\begin{eqnarray}
f'(x_i) = \frac{1}{6\gamma (\delta + x)}
\left ( -11f(x_i)+18f(x_{i+1})-9f(x_{i+2})+2f(x_{i+3}) \right ) \, , \\
f'(x_i) = \frac{1}{6\gamma (\delta + x)} 
\left (-2f(x_{i-1}) - 3f(x_i)+6f(x_{i+1})-f(x_{i+2}) \right ) \, , \\
f'(x_i) = \frac{1}{6\gamma (\delta + x)}
\left (f(x_{i-2}) - 6f(x_{i-1}) + 3f(x_i)+2f(x_{i+1}) \right ) \, , \\
f'(x_i) = \frac{1}{6\gamma (\delta + x)}
\left (-2f(x_{i-3}) + 9f(x_{i-2}) - 18f(x_{i-1})+11f(x_i) \right ) \, .
\end{eqnarray}
For $x<0$ the denominator should be replaced by $6\gamma(\delta -x)$.

To verify the convolution code, a computation was done for
a simple functional form GPDs. For the non-singlet,
spin-independent quark-gluon and gluon-gluon parts and spin-dependent
quark-quark and gluon-quark parts
\begin{eqnarray}
H_{test}(x,\xi,Q^2) = 1 - x^2
\end{eqnarray}
was taken, and for singlet spin-independent quark-quark and gluon-quark parts
and spin-dependent quark-gluon and gluon-gluon parts
\begin{eqnarray}
H_{test}(x,\xi,Q^2) = x - x^3 \, .
\end{eqnarray}
It was found that taking $n=50$ provides 4-6 digits precision
for all the integrals. It should be noted that the
test functions are not convenient for the Simpson method calculation
since in the logarithmic grid they fall steeply when $x\to 1$. However,
they made possible analytical computation of the integrals and therefore
they were used in the test. The real GPDs which fall rapidly as
$x\to \pm 1$ should be computed with use of Simpson's method
more precisely on such a grid.

\section{The evolution procedure}
Basing on the integrals obtained in the previous section,
the code for GPD evolution was written. It is based on the 4-th
order Runge-Kutta method. Indeed, after introduction of the
grid (\ref{lngrid}) the convolution integrals are transformed
into sums. Therefore, the integro-differential evolution equation
is transformed into a set of $4n+1$ ordinary differential
equations of the form\footnote{In singlet case, the function
index runs from 0 to $2n$ instead of 0 to $4n$ in the non-singlet
sector.}
\begin{equation}
\frac{d\,H_i}{d\ln Q^2}=F_i(\ln Q^2,H_0, H_1, ... H_{4n}),
\end{equation}
where $H_i = H(x_i,\xi,t,\ln Q)$. The 4-th order Runge-Kutta scheme
on the $\ln Q^2$-grid with $m$ nodes
\begin{equation}
\ln Q^2_j =\ln Q^2_0 + j d\,\ln Q^2\, , \quad 0\le j \le m-1
\label{grid_Q2}
\end{equation}
therefore reads:
\begin{equation}
H_i^j = H_i^{j-1} + \frac{d \ln Q^2}{6}\left ( k^{(1)}_i + 2k^{(2)}_i + 
2k^{(3)}_i +k^{(4)}_i \right) ,
\end{equation}
where\footnote{R.h.s. of these equations depend
explicitly (i.e. not via GPDs) on $Q^2$ through $\alpha_s(Q^2)$}
\begin{eqnarray}
k^{(1)}_i = F_i(\ln Q^2_j,H_0, H_1, ... H_{4n}), \\
k^{(2)}_i = F_i(\ln Q^2_j+\frac{d\,\ln Q^2}{2},H_0+\frac{d\,\ln Q^2}{2}k^{(1)}_0,
H_1+\frac{d\,\ln Q^2}{2}k^{(1)}_1, ... H_{4n}+\frac{d\, \ln Q^2}{2}k^{(1)}_{4n},), \\
k^{(3)}_i = F_i(\ln Q^2_j+\frac{d\,\ln Q^2}{2},H_0+\frac{d\,\ln Q^2}{2}k^{(2)}_0,
H_1+\frac{d\,\ln Q^2}{2}k^{(2)}_1, ... H_{4n}+\frac{d\,\ln Q^2}{2}k^{(2)}_{4n},), \\
k^{(4)}_i = F_i(\ln Q^2_j+ d\,\ln Q^2,H_0+d\,\ln Q^2 \, k^{(3)}_0,
H_1+d\,\ln Q^2 \, k^{(3)}_1, ... H_{4n}+d\,\ln Q^2\,k^{(3)}_{4n},) \, .
\end{eqnarray}

\begin{figure}
\centering
\begin{minipage}[c]{0.49\hsize}
\epsfig{file=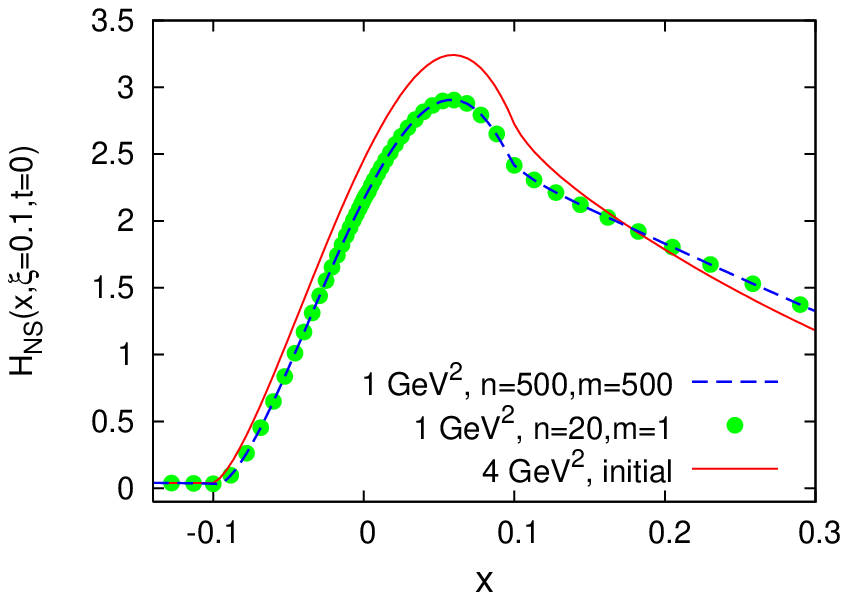,width=\hsize}
\end{minipage}
\begin{minipage}[c]{0.49\hsize}
\epsfig{file=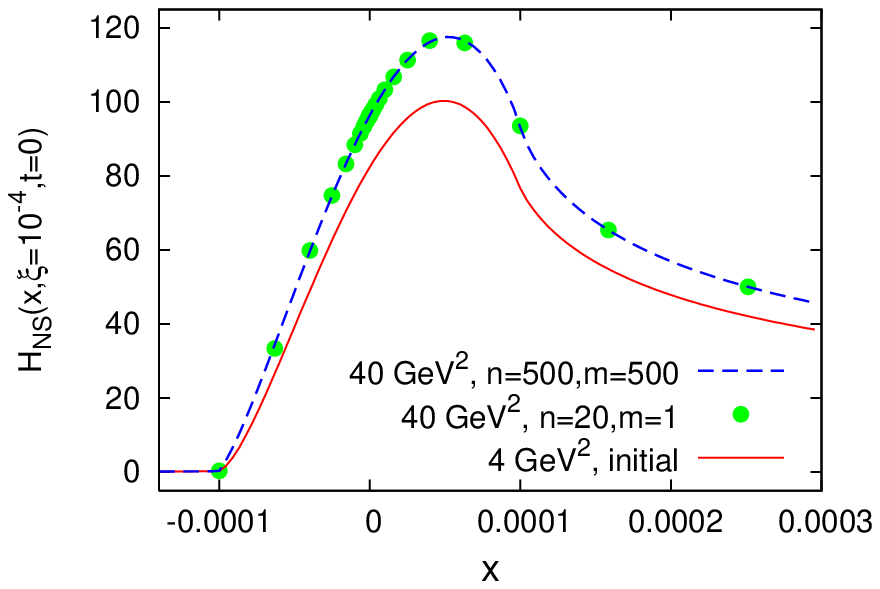,width=\hsize}
\end{minipage}
\caption{Evolution of the quark non-singlet GPD $H_{NS}$ for $\xi = 0.1$
from $Q^2$ = 4 GeV$^2$ to 1 GeV$^2$ (left panel)
and for $\xi=10^{-4}$ from $Q^2$ = 4 GeV$^2$ to 40 GeV$^2$
(right panel). The initial GPD is obtained using
the double-distribution ansatz \cite{radyushkin,musatov} with $b=1$
and CTEQ6M \cite{cteq} PDF at the scale of 4 GeV$^2$. For the evolved
distribution, $n$ and $m$ represent parameters of the
numerical evolution procedure described in Eqs.~(\ref{lngrid}) and
(\ref{grid_Q2}) correspondingly. }
\label{figns}
\end{figure}

\begin{figure}
\centering
\begin{minipage}[c]{0.49\hsize}
\epsfig{file=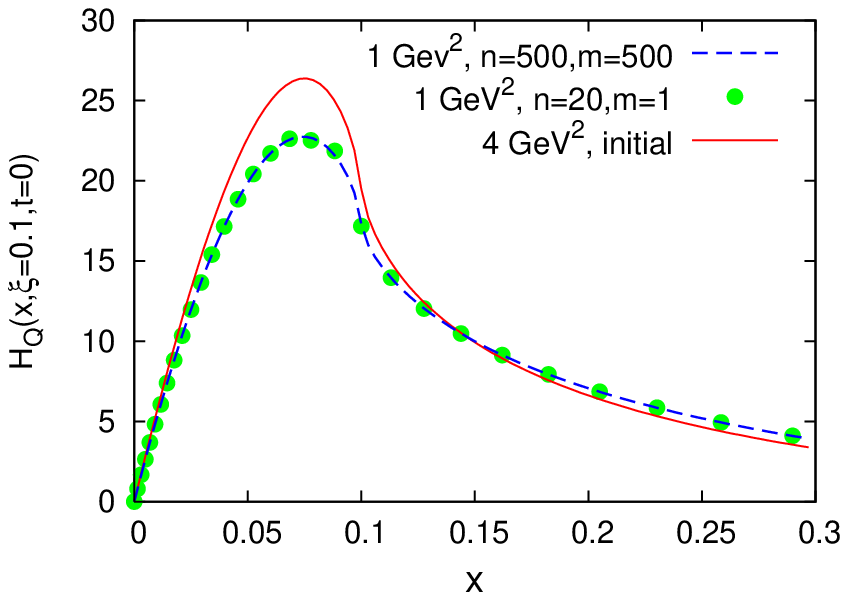,width=\hsize}
\end{minipage}
\begin{minipage}[c]{0.49\hsize}
\epsfig{file=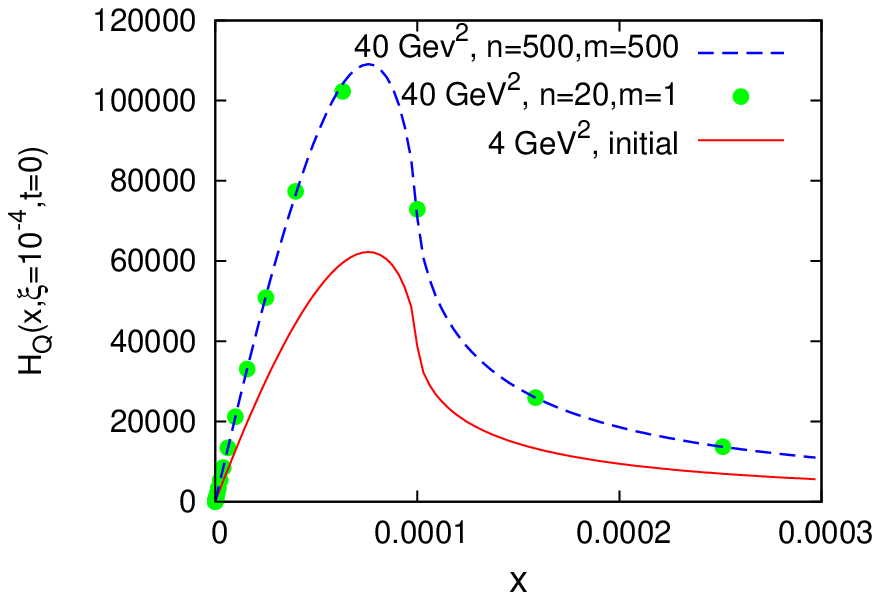,width=\hsize}
\end{minipage}
\caption{Evolution of the quark singlet GPD $H_Q$ for $\xi = 0.1$ 
from $Q^2$ = 4 GeV$^2$ to 1 GeV$^2$ (left panel)
and for $\xi=10^{-4}$ from $Q^2$ = 4 GeV$^2$ to 40 GeV$^2$
(right panel).
See caption of Fig.~\ref{figns}.}
\label{figq}
\end{figure}

\begin{figure}
\centering
\begin{minipage}[c]{0.49\hsize}
\epsfig{file=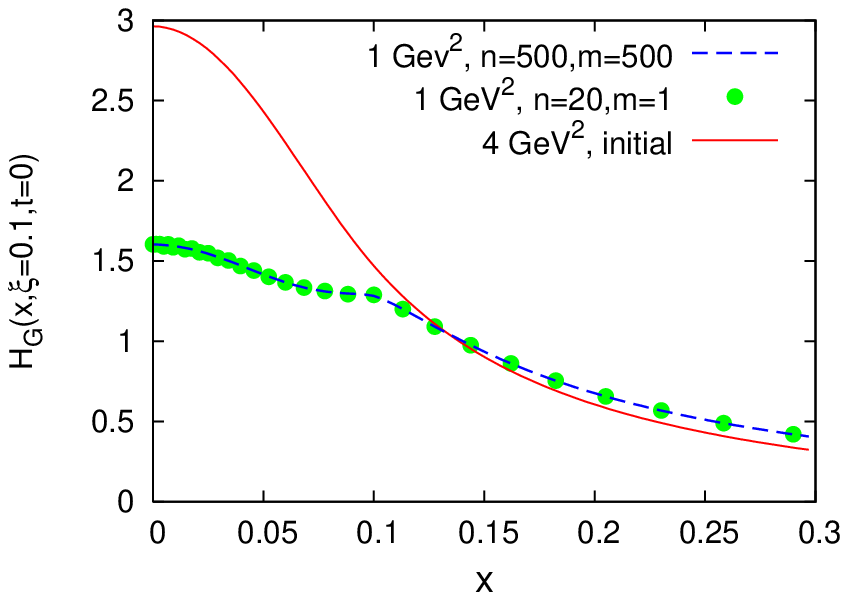,width=\hsize}
\end{minipage}
\begin{minipage}[c]{0.49\hsize}
\epsfig{file=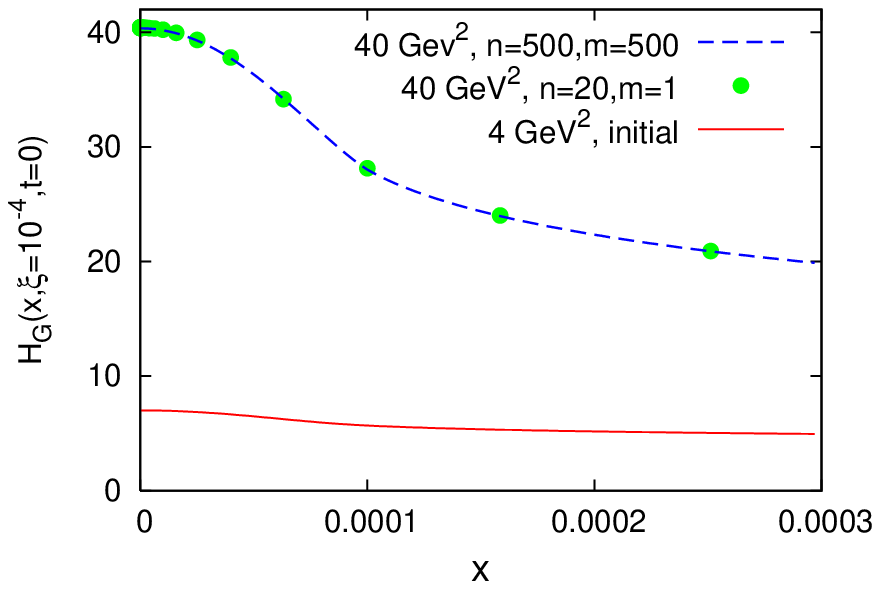,width=\hsize}
\end{minipage}
\caption{Evolution of the gluon GPD $H_G$ for $\xi = 0.1$
from $Q^2$ = 4 GeV$^2$ to 1 GeV$^2$ (left panel)
and for $\xi=10^{-4}$ from $Q^2$ = 4 GeV$^2$ to 40 GeV$^2$
(right panel).
See caption of Fig.~\ref{figns} (the input profile parameter $b=2$
was taken).}
\label{figg}
\end{figure}

\begin{figure}
\centering
\begin{minipage}[c]{0.49\hsize}
\epsfig{file=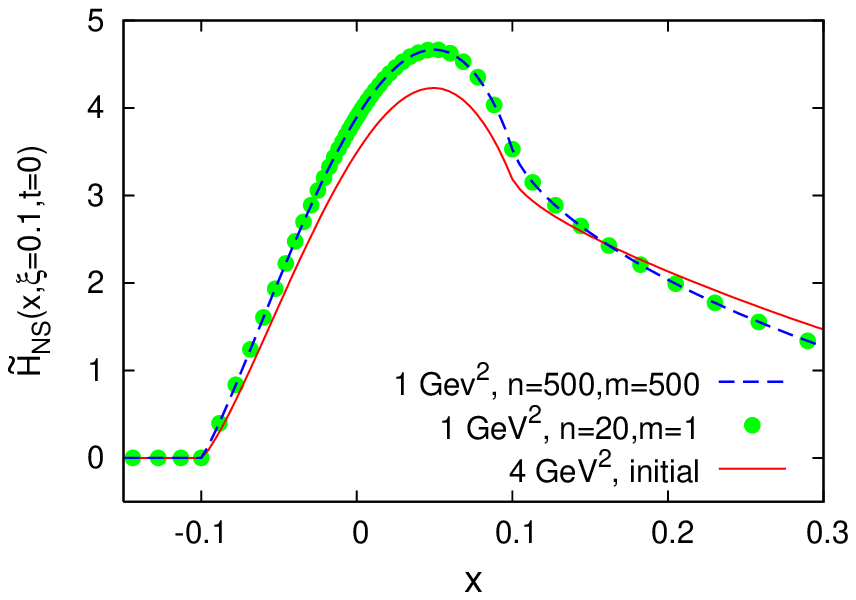,width=\hsize}
\end{minipage}
\begin{minipage}[c]{0.49\hsize}
\epsfig{file=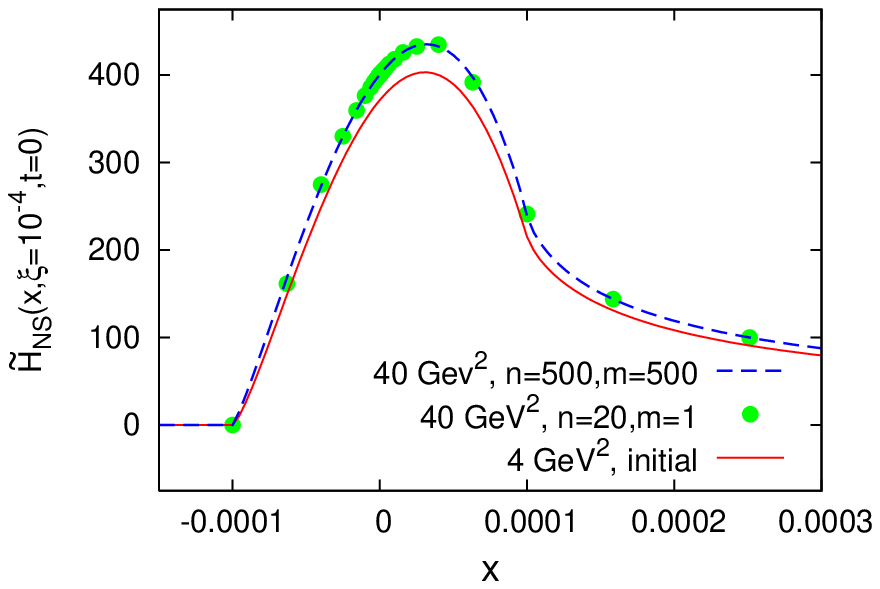,width=\hsize}
\end{minipage}
\caption{Evolution of the spin-dependent non-singlet GPD
$\widetilde H_{NS}$ for $\xi = 0.1$
from $Q^2$ = 4 GeV$^2$ to 1 GeV$^2$ (left panel)
and for $\xi=10^{-4}$ from $Q^2$ = 4 GeV$^2$ to 40 GeV$^2$
(right panel). The initial GPD is obtained using
the double-distribution ansatz \cite{radyushkin,musatov} with $b=1$
and Bl\"umlein-B\"ottcher \cite{blumbott}
PDF at the scale of 4 GeV$^2$. For the evolved
distribution, $n$ and $m$ represent parameters of the
numerical evolution procedure described in Eqs.~\ref{lngrid} and
\ref{grid_Q2} correspondingly. }
\label{figns_pol}
\end{figure}

\begin{figure}
\centering
\begin{minipage}[c]{0.49\hsize}
\epsfig{file=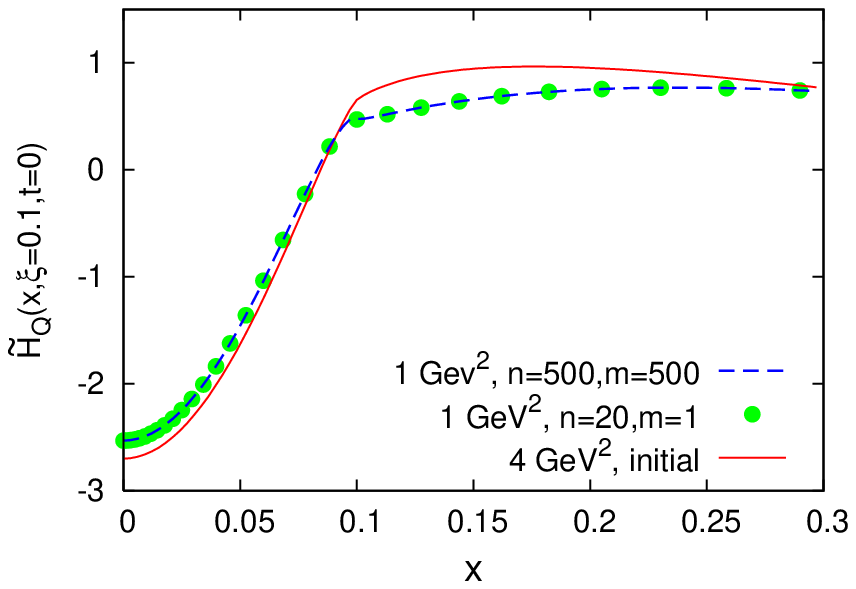,width=\hsize}
\end{minipage}
\begin{minipage}[c]{0.49\hsize}
\epsfig{file=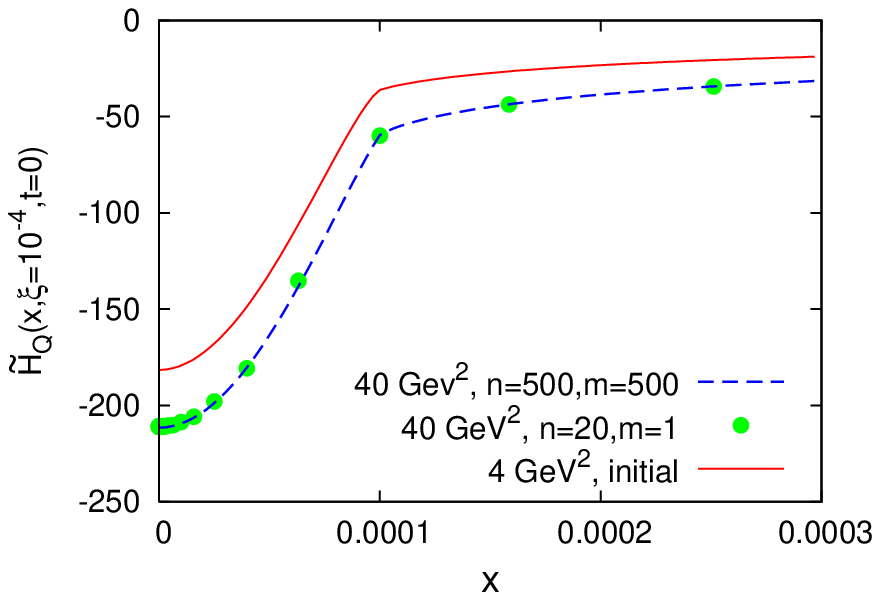,width=\hsize}
\end{minipage}
\vspace{-3mm}
\caption{Evolution of the spin-dependent singlet quark GPD
$\widetilde H_Q$ for $\xi = 0.1$
from $Q^2$ = 4 GeV$^2$ to 1 GeV$^2$ (left panel)
and for $\xi=10^{-4}$ from $Q^2$ = 4 GeV$^2$ to 40 GeV$^2$
(right panel).
See caption of Fig.~\ref{figns_pol}.}
\label{figq_pol}
\end{figure}

\begin{figure}
\centering
\begin{minipage}[c]{0.49\hsize}
\epsfig{file=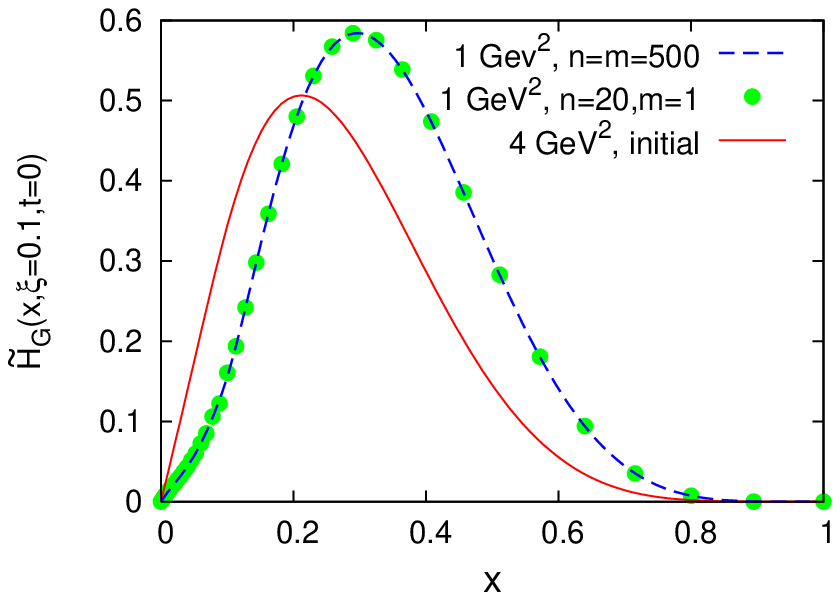,width=\hsize}
\end{minipage}
\begin{minipage}[c]{0.49\hsize}
\epsfig{file=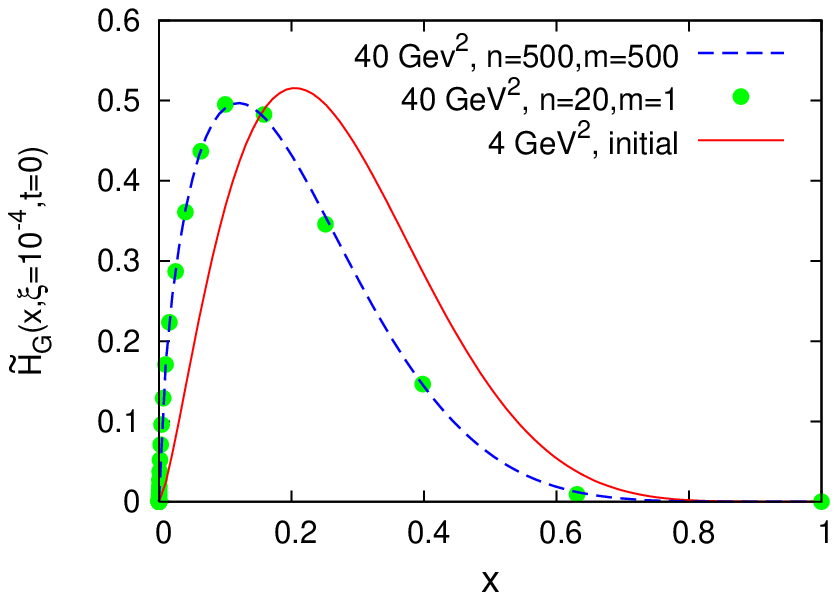,width=\hsize}
\end{minipage}
\vspace{-3mm}
\caption{Evolution of the spin-dependent gluon GPD
$\widetilde H_G$ for $\xi = 0.1$
from $Q^2$ = 4 GeV$^2$ to 1 GeV$^2$ (left panel)
and for $\xi=10^{-4}$ from $Q^2$ = 4 GeV$^2$ to 40 GeV$^2$
(right panel). See caption of Fig.~\ref{figns_pol}
(the input profile parameter $b=2$ was taken).}
\label{figg_pol}
\end{figure}

The results of the computation are presented
in Figs.~\ref{figns}-\ref{figg_pol}
for typical kinematics of HERMES and HERA. The code works
both in forward (from smaller to larger $Q^2$) and backward
(from larger to smaller $Q^2$) directions. As it can be seen,
choice of $n=20$ and $m=1$ already gives satisfactory
results. Distinctions from the careful solution with $n=500$
and $m=500$ can be noticed around $x=\xi$,
but they are much smaller then typical
experimental errors, so the choice of $n=20$, $m=1$ can be recommended
for time-consuming fitting procedures. If the time limitations are not
very strict, larger number of $n$ and $m$ can be taken providing
higher precision.

The evolution procedure was checked for satisfying the momentum
conservation sum rule
\begin{equation}
\frac{d}{d\,\ln Q^2}\int\limits_{-1}^1 \left ( x H_Q(x,\xi,Q^2)
+ H_G(x,\xi,Q^2) \right ) dx \, = \, 0
\end{equation}
and flavor conservation sum rule
\begin{equation}
\frac{d}{d\,\ln Q^2}\int\limits_{-1}^1 H_{NS}(x,\xi,Q^2) dx \, = \, 0 .
\end{equation}
For $n=20$, $m=1$, the sum rules are satisfied within 0.01\% accuracy
at HERMES kinematics and 0.5\% accuracy at HERA kinematics.

Computational time depends on $n$ and $m$ as
\begin{equation}
t=\alpha n^2 m,
\end{equation}
where $m$ is the number of intervals for $\ln Q$. The coefficient
$\alpha$ depends on the computational power of the machine.
For Pentium-IV 1.7 GHz $\alpha\approx 1.4\cdot 10^{-6}$ sec in the non-singlet
case and $\approx$ $2\cdot 10^{-6}$ in the singlet case.
This means that computation of evolution with $n$=20, $m$=1
takes about $10^{-3}$ sec.


\section{Existing numerical instabilities}
\begin{figure}
\centering
\begin{minipage}[c]{0.49\hsize}
\epsfig{file=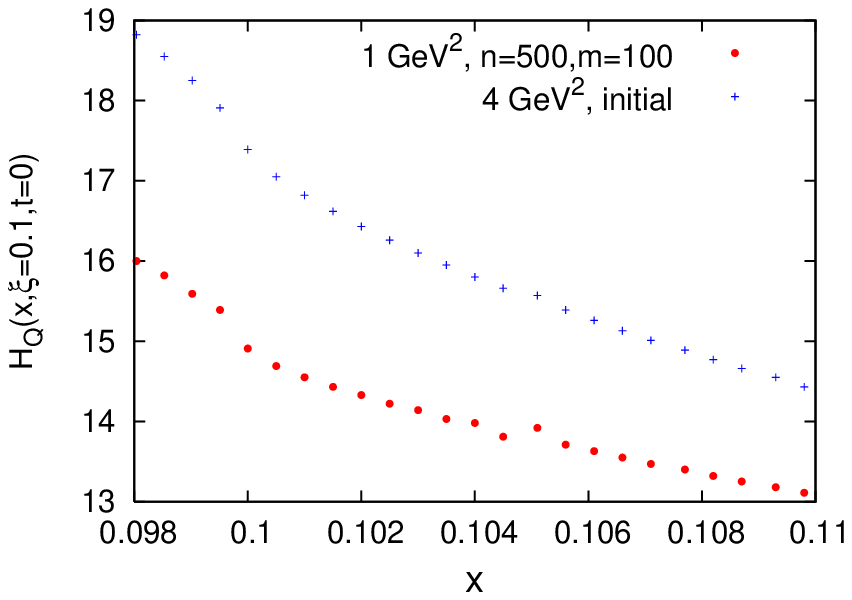,width=\hsize}
\end{minipage}
\begin{minipage}[c]{0.49\hsize}
\epsfig{file=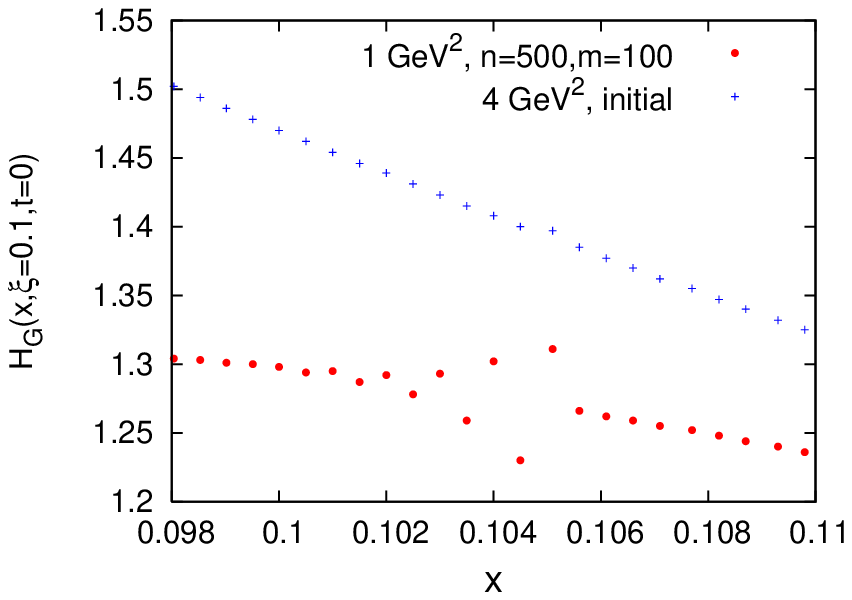,width=\hsize}
\end{minipage}
\caption{Influence of a small (0.003 of the corresponding GPD value)
hand-made spike at $x=0.105$ in the input GPD with $\xi=0.1$
on the output quark (left) and gluon (right) GPDs.}
\label{instab}
\end{figure}

Even though the code is quite effective, instabilities stemming from
the complicate structure of the kernels still arise in the output.
So far, only one instability was found. It is related
to the fact that due to the ``+'' regularization procedure the
convolution integral at a given value of $x$
depends to a large extent on the value of derivative of the GPD at this
point. Therefore, if an input GPD has sudden spikes (not related to
a possible semi-singular behavior at $x=\xi$ which the code can handle),
the code can amplify them. This type of instability is illustrated
in Fig.~\ref{instab}. The problem occurs only in the gluon-gluon part
of the kernel and only for evolution in the backward direction,
{\it i.e.} from larger $Q^2$ to smaller one. It should be
noted that the smaller is the step in $x$, the bigger is the value
of the derivative the sudden spikes produce. Therefore, when for some
reason the task is to perform evolution with very high precision,
the stability of the input GPDs must be taken care of.

\section{Summary}
A code for prompt numerical computation of leading order GPDs evolution
is presented. Its fundamentals are computation of the convolution integrals
on a logarithmic grid and 4-th order Runge-Kutta method for integration
of the differential equations. The code makes possible fast computation
of the GPD evolution. On a 1.7 GHz Pentium-4 machine computational time
required to perform evolution in the $Q^2$ range from 4 to 1 GeV$^2$
or from 4 to 40 GeV$^2$
is of order of 10$^{-3}$ seconds. Such code makes possible fitting of GPDs
parameters from data on hard elastic lepton-nucleon reactions.

\section*{Acknowledgements}
This work from the very beginning
would be impossible without guidance of M.~Diehl.
His advise also made the code more stable and effective.
The author is grateful to W.-D.~Nowak for permanent support
and useful discussions.
The work was supported by the Alexander von Humboldt foundation
and in part by RFBR grant 04-02-16445.

\setcounter{section}{0}
\def\thesection{\Alph{section}}
\section{Explicit expressions for the kernels}

\subsection{Non-singlet part}
After cancellation of the singularities thanks to the ``+'' operation
the kernels are ready for practical usage. 
At $x<-\xi$ the non-singlet part of the evolution equation reads:
\begin{eqnarray}
\frac{d H_{NS}(x,\xi,Q^2)}{d \ln Q^2}\, =\, \frac{2\alpha_s(Q^2)}{3\pi}
\Biggl [ \int\limits_{-1}^x dy
\frac{2\xi^2 - x^2 - y^2}{(y-x)(y^2-\xi^2)}
\left ( H_{NS}(y,\xi,Q^2) - H_{NS}(x,\xi,Q^2) \right ) \nonumber \\
+\quad H_{NS}(x,\xi,Q^2)\, 
\Biggl ( \frac{3}{2} +  2\ln(1+x) +
\frac{x-\xi}{2\xi}\ln\left ( (\xi-x)(1-\xi) \right ) \\
\left. -\frac{x+\xi}{2\xi}\ln\left ( (-\xi-x)(1+\xi) \right )
\Biggr ) \, \right ] \, , \nonumber
\end{eqnarray}
At $-\xi<x<\xi$:
\begin{eqnarray}
\frac{d H_{NS}(x,\xi,Q^2)}{d \ln Q^2} \, = \, \frac{\alpha_s(Q^2)}{3\pi}
\Biggl [ \int\limits_x^1 dy \frac{(x+\xi)(y-x+2\xi)}{\xi(y+\xi)(y-x)}
\left ( H_{NS}(y,\xi,Q^2)-H_{NS}(x,\xi,Q^2)\right ) \nonumber \\
+\, \int\limits_{-1}^x dy \frac{(x-\xi)(y-x-2\xi)}
{\xi(y-\xi)(y-x)}\left ( H_{NS}(y,\xi,Q^2)-H_{NS}(x,\xi,Q^2) \right ) \\
+\, H_{NS}(x,\xi,Q^2)\left (3+2\ln\frac{1-x^2}{1+\xi} +
\frac{x-\xi}{\xi}
\ln(\xi-x) -\frac{x+\xi}{\xi}\ln(x+\xi) \right ) \, \Biggr ] \, ,
\nonumber
\end{eqnarray}
$x > \xi$:
\begin{eqnarray}
\frac{d H_{NS}(x,\xi,Q^2)}{d \ln Q^2} \, = \, \frac{2\alpha_s(Q^2)}{3\pi}
\Biggl [ \int\limits_x^1 dy
\frac{x^2+y^2-2\xi^2}{(y-x)(y^2-\xi^2)}
\left ( H_{NS}(y,\xi,Q^2) - H_{NS}(x,\xi,Q^2) \right ) \nonumber \\
+\, H_{NS}(x,\xi,Q^2)
\Biggl ( \frac{3}{2} + 2\ln(1-x) +\frac{x-\xi}{2\xi}
\ln\left ( (x-\xi)(1+\xi) \right ) \label{qns3}\\
\left. -\, \frac{x+\xi}{2\xi}\ln\left ( (x+\xi)(1-\xi) \right )
\Biggl ) \quad \right ] . \nonumber
\end{eqnarray}

\subsection{Singlet quark-quark part}
In the DGLAP region singlet quark-quark evolution is governed by the
same kernel as the non-singlet (Eq~.\ref{qns3}). At $0< x<\xi$
\begin{eqnarray}
\frac{d_Q H_Q(x,\xi,Q^2)}{d\ln Q^2} = \frac{\alpha_s(Q^2)}{3\pi}
\Biggl [ \frac{2x}{\xi}\int\limits_x^1 \frac{H_Q(y,\xi,Q^2)}{\xi+y}dy
+4x\int\limits_x^1\frac{H_Q(y,\xi,Q^2)-H_Q(x,\xi,Q^2)}{y^2-x^2}dy \nonumber \\
+ 2\frac{\xi-x}{\xi}\int\limits_0^x \frac{H_Q(y,\xi,Q^2)}{\xi^2-y^2}ydy
-4(\xi^2-x^2)\int\limits_0^x\frac{H_Q(y,\xi,Q^2)-H_Q(x,\xi,Q^2)}
{(\xi^2-y^2)(y^2-x^2)}ydy \\
+\quad 2H_Q(x,\xi,Q^2)\left ( \ln\frac{1-x}{1+x}
+ 2\ln\frac{x}{\xi} +\frac{3}{2}
\right ) \, \Biggr ] \, . \nonumber
\end{eqnarray}

\subsection{Quark-gluon part}
In this part and also in the gluon-gluon part of the kernel
a pole at $y=\xi$ must be regularized. For this, subtraction
of the $H_G(\xi,\xi,Q^2)$ is done. If $H_G(x,\xi,Q^2)$ is subtracted
instead, this lead to larger of inaccuracy of the Simpson's method
since the function
$$
\left ( H_G(y,\xi,Q^2) - H_G(x,\xi,Q^2) \right )/(y-\xi)
$$
varies very rapidly around $y=x$.

At $0<x<\xi$:
\begin{eqnarray}
\frac{d_G H_Q(x,\xi,Q^2)}{d\ln Q^2} = 
\qquad\qquad\qquad\qquad\qquad\qquad\qquad\qquad\qquad\qquad\qquad
\qquad\qquad\qquad \nonumber \\
n_f\frac{\alpha_s(Q^2)}{2\pi\xi^2}
\Biggl [ \int\limits_0^x\frac{\xi^4 + \xi^3 x - 2\xi^2 x^2 -
(\xi^2 - \xi x)y^2}{(\xi^2-y^2)^2}\left ( H_G(y,\xi,Q^2) - H_G(\xi,\xi,Q^2)
\right ) dy \nonumber \\
+\xi x\int\limits_x^1\frac{H_G(y,\xi,Q^2)}{(\xi+y)^2}dy
+ H_G(\xi,\xi,Q^2)\left ( \frac{x^2}{\xi + x} + \frac{\xi^2 - x^2}{2\xi}
\ln\frac{\xi+x}{\xi-x} \right ) \, \Biggr ] \,
\end{eqnarray}
$x> \xi$:
\begin{eqnarray}
\frac{d_G H_Q(x,\xi,Q^2)}{d\ln Q^2} = 
n_f\frac{\alpha_s(Q^2)}{2\pi}
\Biggl [ \int\limits_x^1\frac{y^2 - 2xy + 2x^2-\xi^2}
{(\xi^2-y^2)^2}\left ( H_G(y,\xi,Q^2) - H_G(\xi,\xi,Q^2)
\right ) dy \nonumber \\
- H_G(\xi,\xi,Q^2)\left ( \frac{\xi^2-x^2}{2\xi^3}
\ln\frac{(1+\xi)(x-\xi)}{(\xi+x)(1-\xi)}
 + \frac{x - x^2}{\xi^2(\xi^2-1)} \right ) \, \Biggr ] \,
\end{eqnarray}

\subsection{Gluon-quark part}
$0<x<\xi$:
\begin{eqnarray}
\frac{d_Q H_G(x,\xi,Q^2)}{d\ln Q^2} = 
\frac{2\alpha_s(Q^2)}{3\pi}
\Biggl [ \frac{(\xi-x)^2}{\xi}\int\limits_0^x\frac{H_Q(y,\xi,Q^2)
-H_Q(\xi,\xi,Q^2)}{\xi^2-y^2}ydy \\
+\int\limits_x^1\frac{\xi^2 + 2\xi y - x^2}{\xi(\xi+y)}H_Q(y,\xi,Q^2)dy
+H_Q(\xi,\xi,Q^2)\frac{(\xi-x)^2}{2\xi}\ln\frac{\xi^2}{\xi^2-x^2} \Biggr ]\, ,
\nonumber
\end{eqnarray}
$x>\xi$:
\begin{equation}
\frac{d_Q H_G(x,\xi,Q^2)}{d\ln Q^2} = 
\frac{2\alpha_s(Q^2)}{3\pi}
\int\limits_x^1\frac{\xi^2-x^2-2y^2+2xy}{\xi^2-y^2}H_Q(y,\xi,Q^2)dy \, ,
\end{equation}

\subsection{Gluon-gluon part}
For the pure gluonic part it was found that simple subtraction
of the GPD value at $y=x$ does not provide high enough precision.
To overcome the difficulty, a partial fractioning was introduced
so that integrand in each of the integrals had only one pole
(at $y=x$ or $y=\xi$).
This brought formal complication into the evolution kernel.
However, practical numerical calculations became more stable
and precise.

$0 < x < \xi$:
\begin{eqnarray}
\frac{d_G H_G(x,\xi,Q^2)}{d\ln Q^2} = \frac{3\alpha_s (Q^2)}{4\pi}
\Biggl [ \quad \int\limits_x^1\frac{6\xi^2+4\xi y - 2x^2}
{\xi(\xi+y)^2}H_G(y,\xi,Q^2) dy \nonumber \\
+\quad 4\int\limits_x^1\frac{\xi^2 y  + 2\xi x^2 + x^2 y}
{(\xi+y)^2 (y^2-x^2)} \left ( H_G(y,\xi,Q^2) - H_G(x,\xi,Q^2) \right ) dy
\nonumber \\
+\quad \frac{(\xi-x)^2}{\xi}\int\limits_0^x\frac{6\xi^2+4\xi x - 2y^2}
{(\xi^2-y^2)^2} \left ( H_G(y,\xi,Q^2) - H_G(\xi,\xi,Q^2) \right ) dy \\
-\quad 4\frac{\xi-x}{\xi+x}\int\limits_0^x\frac{\xi^2 x + 2\xi y ^2 + xy^2}
{(\xi^2-y^2)^2}\left ( H_G(y,\xi,Q^2) - H_G(\xi,\xi,Q^2) \right ) dy
\nonumber \\
-\quad 4\frac{1}{(\xi+x)^2}\int\limits_0^x\frac{\xi^2 x + 2\xi y ^2 + xy^2}
{\xi^2-y^2}\left ( H_G(y,\xi,Q^2) - H_G(\xi,\xi,Q^2) \right ) dy
\nonumber \\
-\quad 4\frac{1}{(\xi+x)^2}\int\limits_0^x\frac{\xi^2 x + 2\xi y ^2 + xy^2}
{y^2-x^2}\left ( H_G(y,\xi,Q^2) - H_G(x,\xi,Q^2) \right ) dy
\nonumber \\
-\quad 4 \frac{\xi-x}{\xi+x}\left ( H_G(\xi,\xi,Q^2) - H_G(x,\xi,Q^2) \right )
\left ( \frac{x}{\xi-x} - \frac{1}{2}\ln\frac{\xi+x}{\xi-x} \right )
\nonumber \\
-\quad 4 \frac{1}{(\xi+x)^2}\left ( H_G(\xi,\xi,Q^2) - H_G(x,\xi,Q^2) \right )
\left ( \xi(\xi+x)\ln\frac{\xi+x}{\xi-x} - x(2\xi+x) \right )
\nonumber \\
+\quad H_G(\xi,\xi,Q^2) \left ( 2(\xi-x)\frac{x}{\xi^2}
+\frac{1}{\xi^2}(2\xi+x)(\xi-x)^2\ln\frac{\xi+x}{\xi-x} \right ) \nonumber \\
+ \quad 2H_G(x,\xi,Q^2)\left ( \ln\frac{1-x^2}{(1+\xi)^2}
-\frac{2}{\xi+1} + \frac{11}{6}-\frac{n_f}{9} \right ) 
\quad \Biggr ] \, , \nonumber
\end{eqnarray}
$x>\xi $:
\begin{eqnarray}
\frac{d_G H_G(x,\xi,Q^2)}{d\ln Q^2} \, = \, 
\frac{3\alpha_s (Q^2)}{\pi}
\Biggl [
\int\limits_x^1\frac{y\xi + x\xi + xy - x^2}{(\xi-y)(\xi+y)^2}
\left ( H_G(y,\xi,Q^2) - H_G(\xi,\xi,Q^2) \right ) dy \nonumber \\
+\int\limits_x^1\frac{H_G(y,\xi,Q^2) - H_G(x,\xi,Q^2)}{y-x} dy\nonumber \\
-\quad(x-\xi)\int\limits_x^1\frac{x^2 + y^2}{(\xi^2-y^2)^2}
\left ( H_G(y,\xi,Q^2) - H_G(\xi,\xi,Q^2) \right ) dy \\
+\quad\left ( H_G(\xi,\xi,Q^2) - H_G(x,\xi,Q^2) \right )
\frac{2\xi^2x-x^3}{4\xi^3}\ln\frac{(\xi-x)(\xi+1)}{(\xi+x)(\xi-1)} \nonumber\\
-\quad\frac{x^2+\xi^2}{2\xi^2}
\left ( H_G(\xi,\xi,Q^2) - H_G(x,\xi,Q^2) \right )
\left ( \frac{\xi (1-x)}{(\xi+x)(\xi+1)}
+\frac{x-\xi}{\xi^2-1}
+ \frac{x}{\xi+x} \right ) \nonumber \\
+ \quad H_G(x,\xi,Q^2) \Biggl ( \frac{(x^2+\xi^2)(1-x)}{2\xi^2(\xi^2-1)}
+ \left ( \frac{1}{2\xi^2} - \frac{x}{4\xi^3} \right )(\xi+x)^2
\ln\frac{1+\xi}{x+\xi} \nonumber \\
- \frac{1}{2}\ln\frac{1-\xi^2}{(1-x)^2}
+ \left ( \frac{1}{2\xi^2} + \frac{x}{4\xi^3} \right ) (\xi-x)^2
\ln\frac{1-\xi}{x-\xi} + \frac{11}{12}-\frac{n_f}{18} \Biggr ) \, \Biggr ] \, ,
\nonumber
\end{eqnarray}

\subsection{Spin-dependent singlet quark-quark part}
$0<x<\xi$:
\begin{eqnarray}
\frac{d_Q \widetilde H_Q(x,\xi,Q^2)}{d\ln Q^2} = \frac{2\alpha_s(Q^2)}{3\pi}
\, \Biggl [ \, \int\limits_x^1 \frac{2\xi y + x^2 + y^2}{(\xi+y)(y^2-x^2)}
\left ( \widetilde H_Q(y,\xi,Q^2) - \widetilde H_Q(x,\xi,Q^2) \right ) dy
\nonumber \\
-\, (\xi-x)\int\limits_0^x \frac{2\xi x + y^2 + x^2}{(\xi^2 - y^2)(y^2-x^2)}
\left ( \widetilde H_Q(y,\xi,Q^2) - \widetilde H_Q(x,\xi,Q^2) \right ) dy \\
+\, \widetilde H_Q(x,\xi,Q^2)\left ( \ln\frac{1-x^2}{(\xi+x)(\xi+1)}
+ \frac{\xi-x}{2\xi}\ln\frac{\xi+x}{\xi-x} +\frac{3}{2}
\right ) \, \Biggr ] \, . \nonumber
\end{eqnarray}
For $x>\xi$ the evolution kernel is given by the same expression
as the non-singlet spin-independent one (Eq.~\ref{qns3}).

\subsection{Spin-dependent quark-gluon part}
$0<x<\xi$:
\begin{eqnarray}
\frac{d_G \widetilde H_Q(x,\xi,Q^2)}{d\ln Q^2} \quad =
\quad n_f\frac{\alpha_s(Q^2)}{2\pi}
\Biggl [ \, \int\limits_x^1 \frac{\widetilde H_G(y,\xi,Q^2)}{(\xi+y)^2} \, dy
\nonumber \\
-\, 2(\xi-x)\int\limits_0^x
\frac{\widetilde H_G(y,\xi,Q^2)-\widetilde H_G(\xi,\xi,Q^2)}
{(\xi^2-y^2)^2}y\,dy \quad
- \quad \frac{x^2}{\xi^2(\xi+x)} \widetilde H_G(\xi,\xi,Q^2) \, \Biggr ] \, .
\end{eqnarray}
$x>\xi$:
\begin{eqnarray}
\frac{d_G \widetilde H_Q(x,\xi,Q^2)}{d\ln Q^2} = -n_f\frac{\alpha_s(Q^2)}{2\pi}
\Biggl [ \, \int\limits_x^1 \frac{\xi^2 + y^2 - 2xy}{(\xi^2-y^2)^2}
\left ( \widetilde H_G(y,\xi,Q^2)-\widetilde H_G(\xi,\xi,Q^2) \right ) dy \\
+ \, \frac{1+x}{\xi^2-1}\widetilde H_G(\xi,\xi,Q^2) \, \Biggr ] \, . \nonumber
\end{eqnarray}

\subsection{Spin-dependent gluon-quark part}
$0<x<\xi$:
\begin{equation}
\frac{d_Q \widetilde H_G(x,\xi,Q^2)}{d\ln Q^2} = \frac{2\alpha_s(Q^2)}{3\pi}
\left [ 2x\int\limits_x^1 \frac{\widetilde H_Q(y,\xi,Q^2)}{\xi+y} dy
-\frac{(\xi-x)^2}{\xi}\int\limits_0^x
\frac{\widetilde H_Q(y,\xi,Q^2)}{\xi^2-y^2}\, y\,dy
\right ] \, .
\end{equation}
$x>\xi$:
\begin{equation}
\frac{d_Q \widetilde H_G(x,\xi,Q^2)}{d\ln Q^2} = \frac{2\alpha_s(Q^2)}{3\pi}
\int\limits_x^1 \frac{\xi^2 + x^2 - 2xy}{\xi^2-y^2}
\, \widetilde H_Q(y,\xi,Q^2) \, dy \, .
\end{equation}

\subsection{Spin-dependent gluon-gluon part}
As it was in the case of the spin-independent gluon-gluon kernel,
the spin-dependent gluon-gluon kernel requires partial fractioning
of the denominator so that each of the integrals would have only
only one pole, which can be easily handled by subtraction.

$0<x<\xi$:
\begin{eqnarray}
\frac{d_G \widetilde H_G(x,\xi,Q^2)}{d\ln Q^2} = \frac{3\alpha_s(Q^2)}{2\pi}
\Biggl [ \, 2x\int\limits_x^1 \frac{\xi^2 + 2\xi y +2y^2 - x^2}
{(\xi+y)^2 (y^2-x^2)} \left ( \widetilde H_G(y,\xi,Q^2) -
\widetilde H_G(x,\xi,Q^2) \right ) dy
\nonumber \\
+\, 2 \, \frac{\xi-x}{\xi+x} \int\limits_0^x
\frac{y^2 - \xi^2 - 2x^2 -2\xi x}{(\xi^2 - y^2)^2}
\left ( \widetilde H_G(y,\xi,Q^2)-\widetilde H_G(\xi,\xi,Q^2)\right ) \,y\,dy
\nonumber\\
+\, \frac{2}{(\xi+x)^2} \int\limits_0^x
\frac{y^2 - \xi^2 - 2x^2 -2\xi x}{\xi^2 - y^2}
\left ( \widetilde H_G(y,\xi,Q^2)-\widetilde H_G(\xi,\xi,Q^2)\right ) \,y\,dy\\
+\, \frac{2}{(\xi+x)^2} \int\limits_0^x
\frac{y^2 - \xi^2 - 2x^2 -2\xi x}{y^2-x^2}
\left ( \widetilde H_G(y,\xi,Q^2)-\widetilde H_G(x,\xi,Q^2) \right ) \,y\,dy
\nonumber\\
-\, \left ( \widetilde H_G(\xi,\xi,Q^2) - \widetilde H_G(x,\xi,Q^2) \right )
\left ( \frac{2x^3}{\xi^2(\xi+x)} + \frac{x^2}{(\xi+x)^2}
+\ln\frac{\xi^2}{\xi^2-x^2} \right ) \nonumber \\
+\, \widetilde H_G(x,\xi,Q^2)\left ( \frac{2x(1-x)}{(\xi+x)(1+x)}
-\frac{2x^3}{\xi^2 (\xi+x)} + \ln\frac{x^2(1-x)}{\xi^2 (1+x)}
\right ) \,\Biggr ] \, . \nonumber
\end{eqnarray}
$x>\xi$:
\begin{eqnarray}
\frac{d_G \widetilde H_G(x,\xi,Q^2)}{d\ln Q^2} = \frac{3\alpha_s(Q^2)}{2\pi}
\Biggl [ \, 2 \int\limits_x^1
\frac{\widetilde H_G(y,\xi,Q^2)-\widetilde H_G(x,\xi,Q^2)}{y-x} \, dy 
\nonumber\\
+ 2\int\limits_x^1 \frac{x(\xi^2+y^2)-y(x^2+y^2-\xi^2)}
{(\xi^2-y^2)^2}
\left ( \widetilde H_G(y,\xi,Q^2) - \widetilde H_G(\xi,\xi,Q^2) \right )\, dy\\
+\, \widetilde H_G(x,\xi,Q^2)
\left ( \frac{2x-2}{1-\xi^2} + \ln\frac{(1-x)^2}{1-\xi^2} \right )
+ 2\, \widetilde H_G(\xi,\xi,Q^2)
\left ( \frac{2x - \xi^2 - x^2}{\xi^2-1}+1 \right ) \nonumber \\
+ \, \left ( \widetilde H_G(y,\xi,Q^2)-\widetilde H_G(\xi,\xi,Q^2) \right )
\left ( 2\frac{\xi^2-x}{\xi^2-1} - 2 + \ln\frac{\xi^2-x^2}{\xi^2-1} \right )
\,\Biggr ] \, . \nonumber
\end{eqnarray}

\end{document}